\documentclass[aps, prb, twocolumn, amssymb, amsmath, showpacs, superscriptaddress]{revtex4-1}
\usepackage{bm}
\usepackage{times}
\usepackage{graphicx}
\usepackage{makecell}
\usepackage{color}
\usepackage{dcolumn}
\usepackage[colorlinks=true, letterpaper=true, pdfstartview=FitV, linkcolor=blue, citecolor=blue, urlcolor=blue]{hyperref}
\usepackage{appendix}
\usepackage[normalem]{ulem}

\usepackage{braket}

\usepackage{helvet}

\usepackage{pifont}

\begin{document}
\title{Lindbladian Phase Geometry and Hall Transport in Open Bloch Systems}

\author{Zhihao Jiang}
\thanks{These authors contributed equally to this work.}
\affiliation{College of Physics and Optoelectronic Engineering, Shenzhen University, Shenzhen 518060, China}
\author{Longjun Xiang}
\thanks{These authors contributed equally to this work.}
\affiliation{College of Physics and Optoelectronic Engineering, Shenzhen University, Shenzhen 518060, China}
\author{Jian Wang}
\email{jianwang@hku.hk}
\affiliation{College of Physics and Optoelectronic Engineering, Shenzhen University, Shenzhen 518060, China}
\affiliation{Quantum Science Center of Guangdong-Hongkong-Macao Greater Bay Area, Shenzhen 518045, China}
\affiliation{Department of Physics, The University of Hong Kong, Pokfulam Road, Hong Kong, China}

\begin{abstract}
Quantum geometry underlies a wide range of transport phenomena in Bloch systems. How quantum-geometric transport is modified when Bloch electrons are coupled to an environment, however, remains largely unexplored, even though the environment can alter both the electronic state and the physical current operator. Here we formulate dc linear response within a trace-preserving Lindblad kinetic theory by defining the physical velocity as the Liouvillian time derivative of the position operator. This construction reveals an environment-induced contribution to the current vertex whose momentum-space curl generates new Hall responses governed by the gauge-invariant phase geometry encoded by Lindblad jump amplitudes, which characterize electron-environment coupling. To leading order in the dissipative coupling, this geometry gives rise to interband shift-vector and diagonal-vorticity Hall responses arising from off-diagonal and diagonal jump amplitudes, respectively. Remarkably, both mechanisms can produce a finite Hall conductivity even when the conventional Berry-curvature anomalous Hall effect vanishes identically. Our work establishes Lindbladian phase geometry as an independent geometric origin of transverse transport in open quantum matter.
\end{abstract}

\maketitle

\noindent \textit{\textcolor{blue}{Introduction.}}---
Quantum geometry provides a unifying language for transport in crystalline solids. 
In Hermitian Bloch systems, an electric field induces an anomalous velocity governed locally by the Berry curvature, leading to an intrinsic anomalous Hall effect~\cite{Xiao2010,Nagaosa2010}.  Related geometric quantities, including the quantum metric, enter geometric corrections to dissipative transport, nonlinear effects, superfluid weight, optical absorption, shift current, and orbital magnetism~\cite{Sodemann2015,Gao2014,Gao2023, N-Wang2023,Han2024,Wei2023,Xiang2024, Peotta2015,Liang2017,Morimoto2016,Gao2015}. 
These developments establish a powerful closed-system paradigm in which geometric transport is determined by the structure of Hamiltonian Bloch bands together with their occupation.

Open Bloch systems challenge this paradigm. Environmental coupling, relaxation, gain, and loss are not merely imperfections but can be intrinsic or engineered ingredients of quantum materials and synthetic platforms. Such systems are described by effective non-Hermitian Hamiltonians or, when trace preservation is essential, by Lindblad kinetic equations \cite{Ashida2020,Bergholtz2021,Breuer2002,Daley2014}. Environments can generate steady-state coherence, modify topological structures, and induce dynamical phases \cite{Minganti2018,Lieu2019}. Although Hall conductance, response theory, adiabatic transport, and dissipative topological pumping in open systems have been investigated \cite{Shen2014,Talkington2024,Avron2012,Fedorova2020}, dc geometric transport in extended open Bloch systems remains largely unexplored. A central question is whether the environment merely renormalizes Berry-curvature responses through modified steady-state populations or also supplies a geometric current vertex that generates distinct transverse transport channels.

\begin{figure*}[tbp]
\centering
\includegraphics[width=1.0\linewidth]{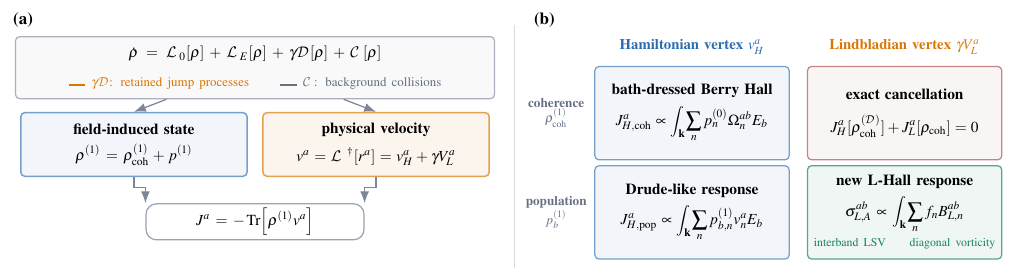}
\caption{DC transport of a Bloch system coupled to a Lindblad environment.
The electric field drives longitudinal and Hall currents, while jump processes
modify both the density matrix and the current vertex through the Lindbladian
velocity.}
\label{fig1}
\end{figure*}

Here we formulate dc linear response in open Bloch systems within the trace-preserving Lindblad framework summarized in Fig.~\ref{fig1}. We define the physical velocity as the Liouvillian time derivative of the position operator, so that it contains both Hamiltonian and Lindbladian contributions. Retaining the Lindbladian contribution, we show that the phases of Lindblad jump amplitudes generate two distinct responses: an interband Lindbladian shift-vector (LSV) response and a diagonal-vorticity response, arising respectively from off-diagonal jump amplitudes and diagonal jump amplitudes with momentum-dependent phases. In the band representation, the band-diagonal and band-off-diagonal density-matrix subspaces define the population and coherence sectors, respectively. The apparent dc current carried by dissipator-generated coherence is cancelled, order by order in the dissipative coupling, by the corresponding off-diagonal Lindbladian-velocity current. By contrast, coherence generated directly by the electric field survives and yields the conventional Berry-curvature Hall response weighted by the bath-modified steady-state population.

For the sector-preserving dissipators considered below, the genuinely new geometric responses arise from contracting the field-induced population with the diagonal Lindbladian velocity. This Lindbladian current vertex produces two distinct Hall channels.
Off-diagonal matrix elements of the jump operators have phases
that combine with the Berry connections of the initial and final
bands to form a Bloch-gauge-invariant shift vector. Weighting this
displacement by the corresponding jump rate produces a contribution
to the diagonal Lindbladian velocity, whose momentum-space curl
generates the interband Lindbladian shift-vector (LSV) Hall response.
Diagonal matrix elements with nontrivial momentum-dependent phases
instead produce a rate-weighted phase velocity, whose momentum-space
curl generates the diagonal-vorticity Hall response. Both channels can remain finite even when the
Hamiltonian Berry curvature vanishes. We isolate the two mechanisms
in real two-band models with vanishing Berry curvature, using
off-diagonal jump amplitudes for the LSV response and diagonal jump amplitudes for the vorticity response. 

\smallskip
\noindent \textit{\textcolor{blue}{Lindbladian phase velocity and Hall response.}}---
We consider a uniform dc electric field in the length gauge and work
at the independent-particle level. 
The density matrix obeys the trace-preserving Lindblad equation
\begin{equation}
\dot{\rho} = \mathcal{L}[\rho]
=
\mathcal{L}_{0}[\rho]
+\mathcal{L}_{E}[\rho]
+\gamma\mathcal{D}[\rho]
+\mathcal{C}[\rho],
\end{equation}
where $\mathcal{L}_{0}[\rho]=-i[H_{0},\rho]$, $\mathcal{L}_{E}[\rho]=-iE_{b}[r^{b},\rho]$, and
\begin{equation}
\mathcal{D}[\rho]=\sum_{\alpha}\left( L_{\alpha}\rho L_{\alpha}^{\dagger} -\frac{1}{2}\{L_{\alpha}^{\dagger}L_{\alpha},\rho\}\right),
\end{equation}
where $L_\alpha$ denotes the jump operator of channel $\alpha$.
Here $\mathcal C$ denotes a phenomenological background collision
operator whose diagonal projections define the population-relaxation
time $\tau_1$ and transport-relaxation time $\tau$. We assume
$\mathcal C^\dagger[r^a]=0$, so that it relaxes the distribution
without providing an additional current vertex~\cite{Talkington2024}.
We set $\hbar=|e|=1$ and use $\gamma$ to track the retained
dissipator strength. Because
$\mathcal L_E^\dagger[r^a]=\mathcal C^\dagger[r^a]=0$, the exact
physical velocity is
\begin{align}
v^{a} &= \mathcal{L}^{\dagger}[r^{a}] = v_{H}^{a}+\gamma V_{L}^{a}, \\
v_{H}^{a} &= \mathcal{L}_{0}^{\dagger}[r^{a}] = i[H_{0},r^{a}],\quad V_{L}^{a} = \mathcal{D}^{\dagger}[r^{a}],
\end{align}
where $v_{H}^{a}$ and $V_{L}^{a}$ are the Hamiltonian and Lindbladian velocities, respectively. 
Thus, $v_H^a$ is the usual closed-system velocity, whereas $V_L^a$ is generated directly by the dissipator. 
The latter is absent in closed systems and represents a bath-generated current vertex.
In the Bloch-band representation, we define the band-covariant derivative $\widetilde{\nabla}_{a}L_{\alpha}=\partial_{k_{a}}L_{\alpha}-i[\mathcal{A}^{a},L_{\alpha}]$, where $\mathcal{A}_{nm}^{a}=i\langle u_{n}|\partial_{k_{a}}u_{m}\rangle$ is the Berry connection with $|u_n\rangle$ the Bloch basis.
Using $[r^{a},L_{\alpha}]=i\widetilde{\nabla}_{a}L_{\alpha}$ gives~\cite{SM}
\begin{equation}
V_{L}^{a} = \frac{i}{2}\sum_{\alpha}\left[L_{\alpha}^{\dagger}\widetilde{\nabla}_{a}L_{\alpha} - (\widetilde{\nabla}_{a}L_{\alpha})^{\dagger}L_{\alpha}\right].
\end{equation}

With the electron-current convention $J^{a}=-\operatorname{Tr}[\rho v^{a}]$, the current separates into $J_H^a=-\operatorname{Tr}[\rho v_H^a]$ and $J_L^a=-\gamma\operatorname{Tr}[\rho V_L^a]$. 
For the stated current-vertex-neutral background, the trace-adjoint
relation between $\mathcal D$ and $\mathcal D^\dagger$ gives an exact cancellation of the current associated with dissipator-generated coherence, order by order in $\gamma$: the Hamiltonian current carried by dissipator-generated coherence cancels the corresponding off-diagonal Lindbladian current, without forcing the full bulk transport current to vanish~\cite{note1}. This holds for
arbitrary retained dissipators, including mixed Lindblad channels.
Consequently, the surviving coherence is generated by $\mathcal{L}_{E}$ acting on the zero-field steady state; a detailed proof is given in the Supplemental Material (SM)~\cite{SM}.

We now specialize to a sector-preserving dissipator, for which the population and coherence sectors decouple; this is realized when every $2\times2$ jump block is either diagonal or off-diagonal. The response for a generic sector-mixing dissipator is given in the SM~\cite{SM}. The coherence cancellation then leaves the field-induced population coupled to $(V_L^a)_{nn}$, and the dc current is
\begin{equation}
J^{a} = J_{H,\mathrm{coh}}^{a} + J_{H,\mathrm{pop}}^{a} + J_{L,\mathrm{pop}}^{a}. \label{eq1}
\end{equation}
Momentum-dependent phases of the Lindblad jump amplitudes may also
generate a field-independent drift; throughout, $J^a$ denotes the
field-induced current with this zero-field contribution subtracted.

We now evaluate Eq.~(\ref{eq1}) for such a dissipator. For the reference
Fermi population $\mathbf f=(f_1,f_2,\ldots)^T$, the projected
relaxation-time approximation gives zero-field population vector
\begin{equation}
{\mathbf p}^{(0)}=(\mathbb I-\gamma\tau_{1} \mathbb K)^{-1}{\mathbf f}.
\end{equation}
Here $\mathbb K$ is the population-rate matrix generated by interband Lindblad transitions, and $\tau_{1}$ describes relaxation toward $\mathbf f$ through degrees of freedom outside the retained Lindblad channels. 
The explicit form of $\mathbb K$  and the detailed derivations for ${\mathbf p}^{(0)}$ and ${\mathbf p}^{(1)}$ are given in SM~\cite{SM}.

Neglecting background dephasing, the electric-field Liouvillian acting on the zero-field steady state generates the field-induced interband coherence, with $f_n-f_m$ replaced by
$p_n^{(0)}-p_m^{(0)}$.
Contracting this coherence with the off-diagonal matrix elements of the Hamiltonian velocity $v_H^a$ gives~\cite{SM}
\begin{equation}
J_{H,\mathrm{coh}}^{a} = \int_{\mathbf{k}}\sum_{n} p_{n}^{(0)}\Omega_{n}^{ab}E_{b},\label{eq2}
\end{equation}
where $\Omega_{n}^{ab}$ is the Berry curvature of band $n$. 
We use the integral convention $\int_{\mathbf{k}}\equiv\int_{\mathrm{BZ}}d^d k/(2\pi)^d$.
This is the open-system counterpart of the intrinsic Berry-curvature Hall effect: the equilibrium occupation $f_n$ is replaced by the bath-renormalized steady-state population $p_n^{(0)}$. 
Within the same relaxation-time approximation, the field-induced diagonal
population is
\begin{equation}
{\mathbf p}_{b}^{(1)}=\tau(\mathbb I-\gamma\tau \mathbb K)^{-1}\partial_{k_{b}}{\mathbf p}^{(0)},
\end{equation}
and couples to the diagonal matrix elements of the Hamiltonian velocity $v_H^a$, generating an open-system Drude-like response~\cite{SM}
\begin{equation}
J_{H,\mathrm{pop}}^{a} = -\int_{\mathbf{k}}\sum_{n} p_{b,n}^{(1)}(v_{H}^{a})_{nn} E_{b},
\ (v_{H}^{a})_{nn} = \partial_{k_{a}}\varepsilon_{n}.
\end{equation}
Here $\tau$ describes momentum relaxation of the field-induced population by degrees of freedom not included in the explicitly retained Lindblad channels.
While $\tau_1$ above controls zero-field repopulation, $\tau$ controls the field-driven population.
For $\gamma=0$, both $J_{H,\mathrm{coh}}^a$ and $J_{H,\mathrm{pop}}^a$ reduce to the Hermitian results.

The field-induced diagonal population couples to the diagonal part of the Lindbladian velocity $(V_L^a)_{nn}$, leading to a genuinely new Lindbladian current vertex admitted in an open system~\cite{SM},
\begin{equation}
J_{L,\mathrm{pop}}^{a} = -\gamma\int_{\mathbf{k}}\sum_{n} p_{b,n}^{(1)}(V_{L}^{a})_{nn}E_{b}.
\end{equation}
Within the projected relaxation-time approximation, the finite-$\gamma$
antisymmetric part of this Lindbladian population response is~\cite{SM}
\begin{align}
\sigma_{L,A}^{ab}(\gamma)
&=
-\frac{\gamma}{2}
\int_{\mathbf k}\sum_n
\left[
p_{b,n}^{(1)}(V_L^a)_{nn}
-
p_{a,n}^{(1)}(V_L^b)_{nn}
\right]
\nonumber \\
&=
\frac{\gamma\tau}{2}
\int_{\mathbf k}\sum_n
f_n \mathcal B_{L,n}^{ab}
+O(\gamma^2),
\end{align}
where
$\mathcal B_{L,n}^{ab}
=\partial_{k_b}(V_L^a)_{nn}
-\partial_{k_a}(V_L^b)_{nn}$.
The first line is exact within the approximation, whereas the second uses
$p_{b,n}^{(1)}=\tau\partial_{k_b}f_n+O(\gamma)$.
This antisymmetric response does not generate Joule heating.
All model results below use the second line; the symmetric response
is given in SM~\cite{SM}.

The diagonal Lindbladian velocity separates into off-diagonal and diagonal jump-channel contributions,
\begin{align}
(V_{L,\mathrm{off}}^{a})_{nn} 
&= \sum_{\alpha,m\neq n}|L_{mn}^{\alpha}|^{2} \left(\mathcal{A}_{m}^{a}-\mathcal{A}_{n}^{a} -\partial_{k_{a}}\phi_{mn}^{\alpha}\right) \label{eq3} \\
(V_{L,\mathrm{diag}}^{a})_{nn}
&= -\sum_{\alpha}|L_{n}^{\alpha}|^{2}\partial_{k_{a}}\phi_{n}^{\alpha}. \label{eq4}
\end{align}
where $L^\alpha_{mn} = \langle u_m|L_\alpha|u_n \rangle$ and $L^\alpha_n \equiv L^\alpha_{nn}$.
The full diagonal element is $(V_L^a)_{nn}=(V_{L,\mathrm{off}}^a)_{nn}+(V_{L,\mathrm{diag}}^a)_{nn}$. 
Their curls generate the interband LSV and diagonal-vorticity Hall responses, respectively. The off-diagonal term has the same gauge-invariant coordinate-shift structure as conventional side jump, but here the displacement is encoded in a Lindblad jump amplitude and enters directly through the Lindbladian current vertex, which is absent from a Hamiltonian-only treatment~\cite{Sinitsyn2006,Hovhannisyan2019}. 
The interpretation of the diagonal Lindbladian velocity as a phase velocity is discussed in the End Matter.

\smallskip
\noindent \textit{\textcolor{blue}{Symmetry and separation of Hall channels.}}---
The Berry-curvature contribution and the Lindbladian-velocity contributions are constrained by symmetries of the full open system, including the Hamiltonian and all jump operators. 
If the full system is time-reversal invariant, its antisymmetric Hall response vanishes.
More generally, time reversal maps a bath texture to its time-reversed counterpart and relates the corresponding response tensors. 
Thus, a time-reversal-breaking bath can generate a Hall response even when the Hamiltonian is time-reversal invariant, with the reciprocal response obtained by reversing the bath texture~\cite{Zhai2023}.

The three geometric Hall channels can be separated by removing their geometric ingredients. 
If $\Omega_{n}^{ab}(\mathbf{k})=0$, the bath-weighted Berry-curvature channel is absent, and the remaining response arises from the
Lindbladian-velocity channels. 
Conversely, if the total $\mathcal B_{L,n}^{ab}=0$, the overall Lindbladian-velocity Hall response
vanishes. 
Setting the diagonal jump amplitudes to zero isolates the interband LSV channel, whereas purely diagonal jump amplitudes isolate the vorticity channel; Hermitian diagonal dephasing has zero vorticity. These conditions separate the bath-weighted Berry-curvature, interband LSV, and diagonal-vorticity Hall responses.
A bath-renormalized Drude-like Hall term can also arise beyond leading order; its relaxation-time scaling is given in the End Matter.

\begin{figure}[tbp]
\centering
\includegraphics[width=0.8\linewidth]{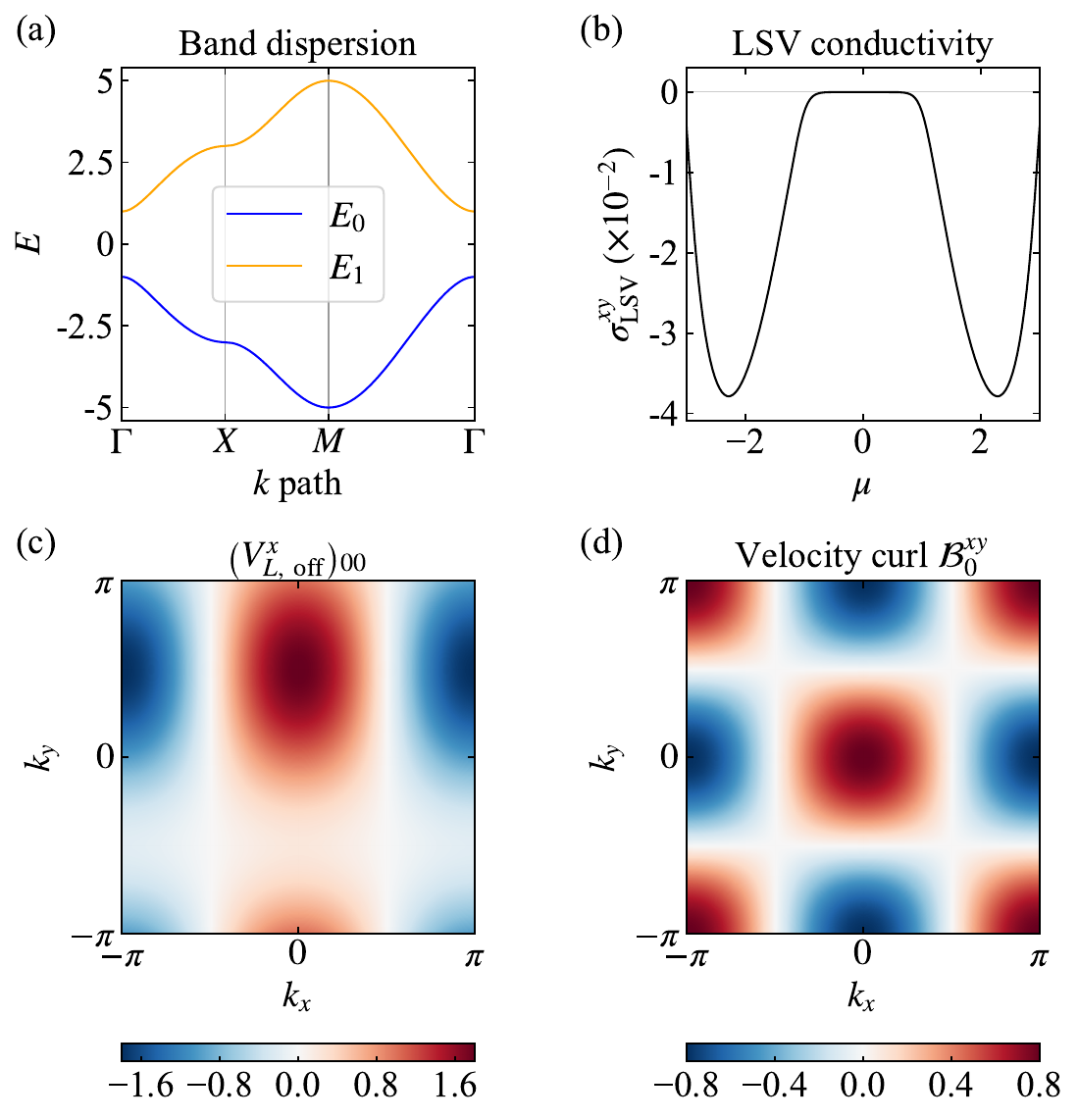}
\caption{Interband LSV Hall effect in the real two-band model.
(a) Band dispersion; (b) antisymmetric LSV conductivity versus chemical
potential; (c) $(V^x_{L,{\rm off}})_{00}(\mathbf k)$; and (d) its curl
$\mathcal B_{L,0}^{xy}(\mathbf k)$.  The Hamiltonian Berry curvature and diagonal-vorticity
channel vanish.  
Since LSV conductivity is proportional to $\gamma\tau$, we choose $\gamma=1,\,\tau=1$. 
Model parameters are $t=m=l_0=\beta=1$, $\eta=0.8$. Temperature is set at $k_BT=0.08$.}
\label{fig2}
\end{figure}

\smallskip
\noindent \textit{\textcolor{blue}{Model calculations.}}---
We illustrate the Lindbladian-velocity-driven Hall response with two minimal models whose conventional Berry-curvature channel vanishes by construction.
We first consider the LSV Hall response for the square-lattice two-band Hamiltonian $H_0(\mathbf{k}) = d_x(\mathbf{k})\sigma_x+d_z(\mathbf{k})\sigma_z$,
with $d_x({\mathbf k})=t\sin k_x$, $d_z({\mathbf k})=m+t(2-\cos k_x-\cos k_y)$.
For $m>0$, the bands are nondegenerate throughout the Brillouin zone.  
Because $H_0$ is real, its Bloch eigenstates can be chosen locally real and $\Omega_n^{xy}(\mathbf{k})=0$ pointwise.  
The bath-weighted Berry-curvature term in Eq.~(\ref{eq2}) therefore vanishes independently of the steady-state population, removing the Berry-curvature Hall channel. 

We realize the LSV response with a single Hermitian off-diagonal Lindblad jump
operator in the band basis,
\begin{align}
 L(\mathbf{k})=\begin{pmatrix}0&L_{01}(\mathbf{k})\\L_{01}^{*}(\mathbf{k})&0\end{pmatrix},   
\end{align}
with $L_{01}({\mathbf k})=\sqrt{l_0(1+\eta\sin k_y)}\,e^{i\beta\sin k_x}$, $|\eta|<1$. 
This is a minimal effective bath texture rather than a unique microscopic model: its lowest harmonics generate a nonzero LSV curvature while preserving zero Hamiltonian Berry curvature and zero diagonal-vorticity response.  
The transition strength $|L_{01}|^2=l_0(1+\eta\sin k_y)$ and phase $\phi_{01}=\beta\sin k_x$ are independently tunable.  
Removing either modulation makes the diagonal Lindbladian velocity curl-free, whereas their variations along $k_y$ and $k_x$, respectively, give $\partial_{k_y}(|L_{01}|^2\partial_{k_x}\phi_{01})\neq0$.  
Experimental routes are discussed below.

In the real gauge of $H_0$, the diagonal Berry connections vanish locally.
The band-resolved LSV therefore follows directly from the phase gradient of
the interband jump amplitude, and its rate-weighted form is
$(V^x_{L,{\rm off}})_{00}=l_0\beta(1+\eta\sin k_y)\cos k_x,
\qquad
(V^y_{L,{\rm off}})_{00}=0$.
Taking its curl gives
opposite LSV curvatures in the two bands.  The curvature and leading
antisymmetric Hall conductivity are
\begin{eqnarray}
\mathcal B_{L,0}^{xy}
&=&\partial_{k_y}(V_{L,{\rm off}}^x)_{00}-\partial_{k_x}(V_{L,{\rm off}}^y)_{00} \nonumber \\
&=&l_0\eta\beta\cos k_x\cos k_y
=-\mathcal B_{L,1}^{xy}, \\
\sigma_{\mathrm{LSV},A}^{xy}
&=&\frac{\gamma\tau}{2}\int_{\mathbf{k}}
\bigl[f_0(\mathbf{k})-f_1(\mathbf{k})\bigr]~\mathcal B_{L,0}^{xy}(\mathbf{k}).
\end{eqnarray}

Figure~\ref{fig2} illustrates these results.  The band structure in panel (a) is fully gapped for $m>0$, consistent with the absence of degeneracies in the real two-band model.  
Panel (c) shows the momentum-space distribution of $(V^x_{L,{\rm off}})_{00}(\mathbf{k})$, whose variation arises from both the transition strength modulation along $k_y$ and the jump-phase gradient along $k_x$.
Taking its curl gives the LSV curvature in panel (d), which changes sign across the Brillouin zone.  
A uniform band occupation therefore gives zero after integration, whereas a metallic Fermi occupation weights different momentum regions unequally.  
Consequently, the LSV Hall conductivity becomes finite when the chemical potential enters a band, as shown in panel (b), providing a direct diagnostic of the LSV mechanism.

The second model defined below isolates the diagonal-vorticity Hall effect via a diagonal jump channel: 
\begin{eqnarray}
\widetilde H_0(\mathbf{k})& =& \widetilde{d}_z(\mathbf{k})\sigma_z, ~~~
\widetilde{d}_z(\mathbf{k})=m+t(2-\cos k_x-\cos k_y),\nonumber\\
\widetilde L(\mathbf{k})
&=&\operatorname{diag}\!\left[\widetilde L_1(\mathbf{k}),
\widetilde L_0(\mathbf{k})\right],~~~\widetilde L_1(\mathbf{k})=\lambda_1,
~~~ |\eta|<1,\nonumber\\
\widetilde L_0(\mathbf{k})
&=&\lambda_0\sqrt{1+\eta\sin k_x}\,e^{i\beta\sin k_y},
\label{eq5}
\end{eqnarray}
With the standard convention $\sigma_z=\operatorname{diag}(1,-1)$, this ordering assigns $\widetilde L_0$ to the lower band $\varepsilon_0=-\widetilde{d}_z$ and $\widetilde L_1$ to the upper band $\varepsilon_1=+\widetilde{d}_z$.
The momentum-independent eigenvectors and the diagonal jump operator eliminate the Hamiltonian Berry-curvature and interband LSV channels, allowing us to isolate the diagonal-vorticity Hall response.

\begin{figure}[t]
\centering
\includegraphics[width=\columnwidth]{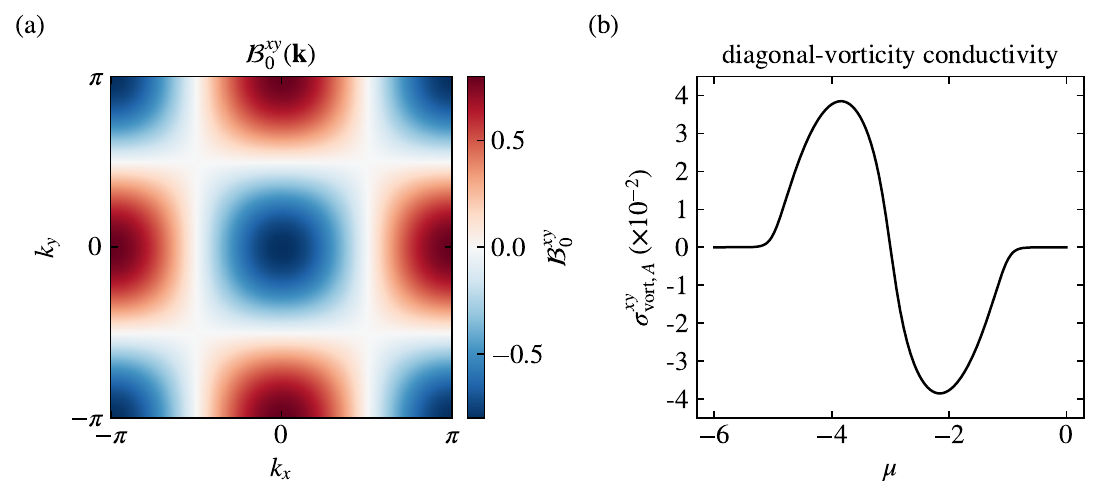}
\caption{Smooth-texture diagonal-vorticity Hall response of the diagonal model in
Eq.~\eqref{eq5}.  (a) Lindbladian vorticity
$\mathcal B_{L,0}^{xy}(\mathbf k)$ and (b) antisymmetric conductivity versus
chemical potential.  The Hamiltonian Berry-curvature and interband LSV
channels vanish.  
Model parameters are $t=m=\lambda_0=\beta=1$, $\eta=0.8$.
As in the LSV case, we set $\gamma=1,\,\tau=1$, and $k_BT=0.08$.}
\label{fig3}
\end{figure}

For the model in Eq.~\eqref{eq5} with a smooth phase texture $\widetilde{\phi}_0=\beta\sin k_y$, the non-vanishing Lindbladian velocity curl comes from nonparallel gradients of the jump strength and phase: $\boldsymbol\nabla_{\mathbf{k}}|\widetilde L_0|^2$ points along $k_x$, whereas $\boldsymbol\nabla_{\mathbf{k}}\widetilde{\phi}_0$ points along $k_y$.
Using Eq.~\eqref{eq4} , we obtain explicitly
\begin{align}
(V^x_{L,{\rm diag}})_{00} &=0,\quad (V_{L,{\rm diag}}^y)_{00} =-\lambda_0^2\beta(1+\eta\sin k_x)\cos k_y,
\nonumber \\
\mathcal B_{L,0}^{xy}(\mathbf{k}) &=\partial_{k_y}(V_{L,{\rm diag}}^x)_{00}-\partial_{k_x}(V_{L,{\rm diag}}^y)_{00} \nonumber \\ &=\lambda_0^2\eta\beta\cos k_x\cos k_y. 
\end{align}
The antisymmetric diagonal-vorticity Hall conductivity is therefore
\begin{equation}
\sigma_{\mathrm{vort},A}^{xy} = \frac{\gamma\tau\lambda_0^2\eta\beta}{2} \int_{\mathbf{k}}f_0(\mathbf{k})\cos k_x\cos k_y. 
\end{equation}
It vanishes if either $\eta$ or $\beta$ is zero.  
Although $\mathcal B_{L,0}^{xy}$ integrates to zero for a uniform occupation, partial occupation weights its positive and negative regions unequally and produces the finite metallic response in Fig.~\ref{fig3}.  
Reversing the phase texture, $\beta\to-\beta$, reverses the Hall signal. 
A singular-phase realization of the vorticity response is given in the End Matter.

\smallskip
\noindent \textit{\textcolor{blue}{Experimental test.}}---
Ultracold atoms in a two-dimensional optical lattice provide a concrete realization, with two long-lived internal or orbital states representing the pseudospin of the real two-band Hamiltonian. 
This realizes the independent-particle regime assumed above.
Reservoir engineering supplies controlled dissipation~\cite{Diehl2008,Damanet2019}, laser-assisted tunneling controls complex hopping
~\cite{Aidelsburger2011,Aidelsburger2013,Miyake2013}, and Raman-driven gases coupled to lossy cavities provide tunable coherent and dissipative couplings~\cite{Ferri2021}. 
Together, these ingredients can realize directional interband dissipative hopping, with the Hermitian-conjugate process included in the same jump channel. The Bloch amplitude ($L_{01}(\mathbf{k})$) is the coherent sum of these directional links, whose independently adjustable magnitudes and phases control the transition-strength and phase textures required for the LSV response.

For the LSV (vorticity) model, a weak force is applied along $y$ ($x$) and the center-of-mass response is measured along $x$ ($y$). Since each minimal model has equal symmetric and antisymmetric transverse parts, the Hall signal is isolated as $\sigma_A^{xy}=(\sigma^{xy}-\sigma^{yx})/2$. The LSV origin is tested by suppressing the momentum dependence of either the transition strength or phase, or by reversing the Raman phase pattern. Suppressing interband transitions and using state-preserving dissipative hopping instead isolates the vorticity response.

Related phase-structured dissipation is available in trapped-ion, photonic, and
circuit-QED platforms~\cite{Poyatos1996,Verstraete2009,Barreiro2011,
Damanet2019,Metelmann2015,Wang2023}; magnetic reservoirs, chiral optical
pumping, and nonreciprocal substrates offer corresponding routes in electronic
or hybrid systems~\cite{Metelmann2015,Wang2023,Silveirinha2025}.

\smallskip
\noindent \textit{\textcolor{blue}{Conclusion.}}---
We formulate dc geometric transport in open Bloch systems at the
independent-particle level using trace-preserving Lindblad kinetics. Defining current from the Liouvillian time
derivative of polarization yields Hamiltonian and Lindbladian velocity terms
and cancels the apparent current from dissipator-generated interband
coherence. 
The field-induced coherence produces the conventional Berry-curvature Hall response dressed by the bath-modified steady state, whereas the population sector coupled with Lindbladian velocity generates new open-system geometric Hall effects.
The latter separates into interband LSV and, for diagonal jump amplitudes, diagonal Lindbladian vorticity responses.
Lindbladian phase geometry thus provides a distinct route to transverse transport in open quantum matter.

\bigskip
We thank the National Natural Science Foundation of China (Grants Nos. 12674063, 12604248, and 12404059).

\newpage
\onecolumngrid
\section*{End Matter}
\par

\twocolumngrid

\smallskip
\noindent \textit{\textcolor{blue}{Lindbladian vorticity.---}}
The terminology can be understood by analogy with spin current vorticity.
Define the spin velocity operator as
\begin{equation}
\hat v^{s,a} =
\frac{1}{2}
\left\{
\hat v^{a},\sigma_s
\right\},
\end{equation}
where $\sigma_s$ is the Pauli matrix for spin component $s$. Within the 
relaxation-time approximation, the field-induced spin current is
\begin{equation}
J^{s,a} 
= \tau \int_{\mathbf{k}} \sum_n E_b \left(\partial_{k_b}f_n\right) v_n^{s,a}
=-\tau\int_{\mathbf{k}}\sum_nE_b f_n \partial_{k_b}v_n^{s,a},
\end{equation}
where
$v_n^{s,a} =\langle u_n|\hat v^{s,a}|u_n\rangle$.
The antisymmetric part of the spin conductivity is therefore
\begin{equation}
\sigma_A^{s,ab}
=
-\frac{\tau}{2}
\int_{\mathbf{k}}
\sum_n
f_n
\left(
\partial_{k_b}v_n^{s,a}
-
\partial_{k_a}v_n^{s,b}
\right).
\end{equation}
The momentum-space curl
\begin{equation}
\omega_n^{s,ab}
=
\partial_{k_b}v_n^{s,a}
-
\partial_{k_a}v_n^{s,b}
\end{equation}
is known as the spin current vorticity~\cite{Mook2020}.
Analogous vorticity structures also occur in layer-polarized responses of twisted systems~\cite{Zhai2023}.

For diagonal non-Hermitian jump operators,
\begin{equation}
(V_{L,\mathrm{diag}}^a)_{nn} = -\sum_{\alpha} \left|L_{n}^{\alpha}\right|^2 \partial_{k_a}\phi_{n}^{\alpha}
\end{equation}
in Eq.~\eqref{eq4} plays the role of a diagonal Lindbladian phase velocity.
To leading order in $\gamma$, its contribution to the antisymmetric conductivity is
\begin{equation}
\sigma_{\mathrm{vort},A}^{ab} =
\frac{\gamma\tau}{2}
\int_{\mathbf{k}}
\sum_n f_n \left[
\partial_{k_b} (V_{L,\mathrm{diag}}^a)_{nn} - \partial_{k_a} (V_{L,\mathrm{diag}}^b)_{nn}
\right].
\end{equation}
Because this response is governed by the momentum-space curl of the diagonal Lindbladian phase velocity, we refer to $\partial_{k_b} (V_{L,\mathrm{diag}}^a)_{nn} - \partial_{k_a} (V_{L,\mathrm{diag}}^b)_{nn}$ as the Lindbladian vorticity and to the resulting transverse response as the diagonal-vorticity Hall effect.

\smallskip
\noindent \textit{\textcolor{blue}{Smooth-texture diagonal-vorticity model.---}}
We now give the details of the second model introduced in Eq.~\eqref{eq5}. 
For $m>0$, we label its two dispersions as $\varepsilon_{0,1}(\mathbf{k})=\mp \widetilde{d}_z(\mathbf{k})$. 
The Bloch eigenvectors are momentum independent, so the diagonal Berry connections and Hamiltonian Berry curvature vanish. 
Moreover, because $\widetilde L$ is diagonal in the band basis, it dephases interband coherence but does not transfer population between the bands. 
Thus the rate matrix is $\mathbb K=0$, and the zero-field and linear field-induced populations reduce to
\begin{equation}
p_n^{(0)}=f_n,
\quad
p_{b,n}^{(1)}=\tau\partial_{k_b}f_n.
\end{equation}
The interband LSV contribution is absent as well.

Writing
$\widetilde L_n=|\widetilde L_n|e^{i\widetilde\phi_n}$, the diagonal
Lindbladian phase velocity is
$(V_{L,\mathrm{diag}}^a)_{nn}=-|\widetilde L_n|^2\partial_{k_a}\widetilde\phi_n$.
For the momentum-dependent lower-band amplitude,
\begin{equation}
|\widetilde L_0|^2=\lambda_0^2(1+\eta\sin k_x),
\qquad
\widetilde\phi_0=\beta\sin k_y,
\end{equation}
whereas $\widetilde L_1=\lambda_1$ has no momentum-dependent phase. Hence
\begin{align}
(V_{L,\mathrm{diag}}^x)_{00}&=0,
&
(V_{L,\mathrm{diag}}^y)_{00}
&=-\lambda_0^2\beta(1+\eta\sin k_x)\cos k_y,
\nonumber\\
(V_{L,\mathrm{diag}}^x)_{11}&=0,
&
(V_{L,\mathrm{diag}}^y)_{11}&=0.
\end{align}
Taking the momentum-space curl gives
\begin{align}
\mathcal B_{L,0}^{xy}(\mathbf{k})
&=\partial_{k_y}(V_{L,\mathrm{diag}}^x)_{00}-\partial_{k_x}(V_{L,\mathrm{diag}}^y)_{00}
\nonumber\\
&=\lambda_0^2\eta\beta\cos k_x\cos k_y,
\qquad
\mathcal B_{L,1}^{xy}(\mathbf{k})=0.
\end{align}
The leading antisymmetric conductivity is therefore
\begin{equation}
\sigma_{\mathrm{vort},A}^{xy}
=\frac{\gamma\tau\lambda_0^2\eta\beta}{2}
\int_{\mathbf{k}}f_0(\mathbf{k})\cos k_x\cos k_y.
\end{equation}

This response is entirely due to the nonparallel momentum-space gradients of
the jump strength and phase:
$\boldsymbol\nabla_{\mathbf{k}}|\widetilde L_0|^2$ points along $k_x$, while
$\boldsymbol\nabla_{\mathbf{k}}\widetilde\phi_0$ points along $k_y$. It
vanishes when either $\eta=0$ or $\beta=0$. Although
$\mathcal B_{L,0}^{xy}$ integrates to zero for a uniform occupation, a
partially occupied band weights its positive and negative regions unequally,
producing the finite metallic response shown in Fig.~\ref{fig3}. Reversing the phase
texture, $\beta\to-\beta$, reverses the Hall signal.

\smallskip
\noindent \textit{\textcolor{blue}{Singular-texture diagonal-vorticity Hall effect.---}}
For a diagonal jump amplitude $L_n^\alpha=|L_n^\alpha|e^{i\phi_n^\alpha}$, the diagonal Lindbladian velocity is
\begin{equation}
(V_{L,\mathrm{diag}}^a)_{nn} = -\sum_\alpha |L_n^\alpha|^2 \partial_{k_a}\phi_n^\alpha .
\end{equation}
Its momentum-space curl can be separated into a smooth amplitude--phase contribution and a phase vortex contribution:
\begin{align}
\mathcal B_{L,n}^{ab}
={}&
    \sum_\alpha
    \left[
        (\partial_{k_a}|L_n^\alpha|^2)
        (\partial_{k_b}\phi_n^\alpha)
        -
        (\partial_{k_b}|L_n^\alpha|^2)
        (\partial_{k_a}\phi_n^\alpha)
    \right]
    \nonumber\\
    &+
    \sum_\alpha
    |L_n^\alpha|^2
    \left(
        \partial_{k_a}\partial_{k_b}
        -
        \partial_{k_b}\partial_{k_a}
    \right)
    \phi_n^\alpha .
    \label{eq6}
\end{align}
For a globally smooth phase, the second line vanishes and Eq.~\eqref{eq6} reduces to the mechanism illustrated in the main text.
A phase vortex, by contrast, can give a distributional contribution through the noncommuting derivatives in the second line.

To isolate this singular contribution, we consider a local continuum model around the $\Gamma$ point whose Hamiltonian is
\begin{equation}
H_\Gamma({\mathbf q}) = d_z({\mathbf q})\sigma_z,
\quad
d_z({\mathbf q}) = m+\frac{t}{2}q^2 ,
\end{equation}
where ${\mathbf q}$ is measured from $\Gamma$.
We introduce the diagonal non-Hermitian jump operator
\begin{equation}
L_\Gamma({\mathbf q}) =
\begin{pmatrix}
    0 & 0\\
    0 & L_0({\mathbf q})
\end{pmatrix},
\quad
L_0({\mathbf q}) = \frac{\lambda_0 (q_x+iq_y)}{\sqrt{q^2+\kappa^2}}.
\end{equation}
The nonzero entry is associated with the lower energy band $\varepsilon_0=-|d_z|$.
Here $\kappa$ regularizes the vortex core.
The jump strength and phase are
\begin{equation}
|L_0({\mathbf q})|^2 = \frac{ \lambda_0^2 q^2}{q^2+\kappa^2},
\quad
\phi_0({\mathbf q}) = \arg(q_x+iq_y).
\end{equation}
The phase winds once around ${\mathbf q}=0$.
The corresponding phase velocity is
\begin{equation}
(V_{{L,\mathrm{diag}}}^{x})_{00} =  \frac{\lambda_0^2 q_y}{q^2+\kappa^2},
\quad
(V_{{L,\mathrm{diag}}}^{y})_{00} = - \frac{\lambda_0^2 q_x}{q^2+\kappa^2}.
\end{equation}
The curl is given by
\begin{equation}
\mathcal B_{L,0}^{xy}({\mathbf q})
= \frac{2\lambda_0^2\kappa^2}{(q^2+\kappa^2)^2}
\xrightarrow{\kappa\rightarrow0} 2\pi\lambda_0^2 \delta^{(2)}({\mathbf q}). 
\end{equation}
Therefore, in the limit $\kappa\rightarrow0$, the singular vortex contribution to the Hall conductivity is
\begin{align}
\sigma_{\mathrm{vort},A}^{xy}
&= \frac{\gamma\tau\lambda_0^2}{4\pi} f_0(\varepsilon_{0,{\mathbf q}=\bm{0}};\mu) \nonumber\\
&= \frac{\gamma\tau\lambda_0^2}{4\pi}
\left[ 1+\exp\left(\frac{-m-\mu}{k_BT}\right)\right]^{-1}.
\end{align}
The response is determined by the occupation at the vortex core.
The sign is reversed for a vortex with opposite winding.

\smallskip
\noindent \textit{\textcolor{blue}{Drude-like population response and relaxation-time
power counting.---}}
The open-system Drude-like population current is given by
\begin{equation}
J_{H,\mathrm{pop}}^{a} = -\int_{{\mathbf k}} \sum_n p_{b,n}^{(1)} v_n^a E_b,\quad v_n^a = \partial_{k_a}\varepsilon_n.
\end{equation}
Within the projected relaxation-time approximation, the two population
formulas stated in the main text are the exact solutions of the
stationary balance equations
\begin{equation}
0=\gamma \mathbb K\mathbf p^{(0)}
-\frac{\mathbf p^{(0)}-\mathbf f}{\tau_1},
~~~
0=\partial_{k_b}\mathbf p^{(0)}
+\gamma \mathbb K \mathbf p_b^{(1)}
-\frac{\mathbf p_b^{(1)}}{\tau}.
\label{eq:population-balance}
\end{equation}
Here $\tau_1$ relaxes the zero-field population toward $\mathbf f$,
whereas $\tau$ relaxes the field-induced transport population.

For relaxation-time power counting, we further expand
$\mathbf p_b^{(1)}
=\sum_{q=0}^{\infty}\gamma^q\mathbf p_b^{(1;q)}$ and
$\mathbf p^{(0)}
=\sum_{s=0}^{\infty}\gamma^s\tau_1^s \mathbb K^s\mathbf f$.
The closed-form inverses require only nonsingularity, whereas these
Neumann series additionally require the spectral radii of
$\gamma\tau \mathbb K$ and $\gamma\tau_1 \mathbb K$ to be smaller than unity.
The coefficient at order $\gamma^q$ is
\begin{equation}
\mathbf p_b^{(1;q)} = \tau \sum_{s=0}^{q} \tau^{q-s}\tau_1^s \mathbb K^{q-s} \partial_{k_b}\!\left(\mathbb K^s \mathbf f\right).
\label{eq7}
\end{equation}
At zeroth order in the dissipative coupling, $p_{b,n}^{(1;0)} = \tau\partial_{k_b} f_n$, and $J^a_{H,\mathrm{pop}}$ reduces to the conventional Drude contribution
\begin{equation}
J_{H,\mathrm{pop}}^{a,(0)} = -\tau \int_{\mathbf k} \sum_n (\partial_{k_b}f_n) v_n^a E_b. 
\end{equation}
The corresponding conductivity is
\begin{equation}
\sigma_{H,\mathrm{pop}}^{ab,(0)} = \tau \int_{\mathbf k} \sum_n \left(-\frac{\partial f}{\partial \varepsilon_n}\right) v_n^av_n^b,
\end{equation}
which is symmetric in $a$ and $b$ and therefore does not contribute to the antisymmetric Hall response.
At first order in $\gamma$, the field-induced population is
\begin{align}
\mathbf p_b^{(1;1)} = \tau^2 \mathbb K\partial_{k_b} \mathbf f + \tau \tau_1 \partial_{k_b}(\mathbb K \mathbf f) .
\end{align}
These two terms have distinct kinetic origins.
The first describes Lindblad redistribution of the ordinary field-induced population, whereas the second describes field driving of the bath-renormalized zero-field population.
The first order Drude-like current response is therefore
\begin{equation}
J_{H,\mathrm{pop}}^{a,(1)} = -\gamma \int_{\mathbf k} \sum_n
\left[ \tau^2 \mathbb K\partial_{k_b} \mathbf f + \tau\tau_1\partial_{k_b}(\mathbb K \mathbf f) \right]_n v_n^a E_b . 
\end{equation}
Unlike the zeroth-order Drude conductivity, the first-order term need not be symmetric for a general momentum- and band-dependent rate matrix $\mathbb K$.

For power counting, assume
$\tau\sim\tau_1\sim\tau_r$. Equation~(\ref{eq7}) then implies
\begin{equation}
p_b^{(1;q)} \sim \tau_r^{q+1},
\quad
J_{H,\mathrm{pop}}^{a,(q)} \sim \gamma^q\tau_r^{q+1}\textcolor{red}{.}
\end{equation}
By comparison, the coherence contribution to the Hamiltonian current at the same order is generated by the qth-order zero-field population correction,
\begin{equation}
\mathbf p^{(0;q)} = \tau_1^q \mathbb K^q \mathbf f,
\quad
J_{H,\mathrm{coh}}^{a,(q)} \sim \gamma^q\tau_r^q\textcolor{red}{.}
\end{equation}
For $q\geq1$, the population current generated by the Lindbladian velocity contains an explicit factor of $\gamma$ and the population response at order $\gamma^{q-1}$,
\begin{equation}
J_{L,\mathrm{pop}}^{a,(q)} = -\gamma^q \int_{\mathbf k} \sum_n p_{b,n}^{(1;q-1)} (V_{L}^a)_{nn} E_b
\sim \gamma^q\tau_r^q.
\end{equation}
Thus, at a fixed order $\gamma^q$, the Drude-like correction contains one additional power of the relaxation time:
\begin{equation}
J_{H,\mathrm{pop}}^{(q)} \sim\gamma^q\tau_r^{q+1},
\quad
J_{H,\mathrm{coh}}^{(q)}, J_{L,\mathrm{pop}}^{(q)} \sim \gamma^q \tau_r^q.
\end{equation}
This power counting is conditional on the projected relaxation-time approximation with $\tau_1$ and $\tau$ and is not universal.

\begin{thebibliography}{99}
\bibitem{note1}
If $\hat r^{a}$ were a bounded observable, stationarity would imply $\operatorname{Tr}[\rho_{\rm ss}\mathcal L^\dagger(\hat r^{a})]=\operatorname{Tr}[(\mathcal L\rho_{\rm ss})\hat r^{a}]=0$. In a periodic bulk, however, position is not a bounded periodic operator; the transport current is instead a flux obtained from the $\mathbf q\rightarrow0$ limit of the finite-$\mathbf q$ density vertex. The adjoint argument above involves only the relevant finite interband position matrix elements and therefore cancels only the coherence contribution.
\end{thebibliography}
\end{document}